\documentclass[prl,twocolumn,showpacs]{revtex4}
\usepackage[dvipdfm]{graphicx}
\usepackage{amsmath,amssymb}
\newcommand{\be}{\begin{equation}}
\newcommand{\ee}{\end{equation}}
\newcommand{\U}{\widetilde{U}}

\begin{document}
\title{Semiclassical structure of chaotic resonance eigenfunctions}
\author{J.P.~Keating$^1$}
\author{M.~Novaes$^1$}
\author{S.D.~Prado$^2$}
\author{M.~Sieber$^1$}
\affiliation{$^1$School of Mathematics, University of Bristol,
Bristol BS8 1TW, UK\\$^2$Instituto de F\'isica, Universidade Federal
do Rio Grande do Sul, 91501-970 Porto Alegre, RS, Brazil}

\begin{abstract}
We study the resonance (or Gamow) eigenstates  of open chaotic
systems in the semiclassical limit, distinguishing between left and
right eigenstates of the non-unitary quantum propagator, and also
between short-lived and long-lived states. The long-lived left
(right) eigenstates are shown to concentrate as $\hbar\to 0$ on the
forward (backward) trapped set of the classical dynamics. The limit
of a sequence of eigenstates $\{\psi(\hbar)\}_{\hbar\to 0}$ is found
to exhibit a remarkably rich structure in phase space that depends
on the corresponding limiting decay rate. These results are
illustrated for the open baker map, for which the probability
density in position space is observed to have self-similarity
properties.
\end{abstract}


\maketitle

In closed systems the two most fundamental semiclassical properties
are that the mean density of states is given by the Weyl law
\cite{weyl}, which associates to each quantum state a Planck cell in
the available region of phase space, and that in classically chaotic
systems the stationary states have Wigner functions which are
semiclassically uniform over the energy shell \cite{berry}, in
agreement with the quantum ergodicity theorem \cite{ergod}. It is
remarkable that it is still not known in general how these
fundamental properties extend to open (scattering) systems.

In open systems the lack of unitarity of the quantum propagator
gives rise to non-orthogonal decaying eigenstates with complex
energies (resonances), the imaginary parts of which are interpreted
as decay rates. In the case of open chaotic systems the classical
mechanics is structured in phase space around fractal sets
associated with trajectories that remain trapped for infinite times,
either in the future (forward-trapped set $K_+$) or in the past
(backward-trapped set $K_-$). The mean density of resonances is
believed (but not in general proved) to be determined by the fractal
dimension of the invariant set $K_0=K_+\cap K_-$, the classical
repeller. This is the fractal Weyl law \cite{frac1,frac2,frac3}.
(Note that this is different to the resonance statistics in weakly
open systems, for which the size of the opening vanishes in the
semiclassical limit ~\cite{Yan}).

Much less is known about the resonance (or Gamow) eigenstates. These
are important in many areas of physics \cite{phys} and chemistry
\cite{chem}, because they have marked influence on observable
quantities such as scattering cross sections and reaction rates
(they are a component of the Siegert pseudostates basis in terms of
which the scattering wavefunctions and S matrix, for example, can
conveniently be expanded \cite{phys}). Following the well
established idea that in the semiclassical limit time-independent
quantum properties should be related to time-independent classical
sets, it is natural to expect that long-lived eigenstates of open
systems should be determined by the structure of $K_+$ and $K_-$.
This was tested numerically for some right eigenstates of the open
kicked rotator in \cite{casati}, where the term `quantum fractal
eigenstates' was coined.

We here significantly extend the notion of quantum fractal
eigenstates in several new directions. First, we draw the important
distinctions between left and right eigenstates of the non-unitary
propagator, and between states that are `short-lived' and
`long-lived' with respect to the Ehrenfest time. Second, we show
that in the semiclassical limit the long-lived left eigenstates
concentrate on $K_+$, while the long-lived right eigenstates
concentrate on $K_-$. In chaotic systems the eigenstates thus
inherit the intricate fractal structure of the underlying classical
trapped sets (this property has also been observed by Nonnenmacher
and Rubin \cite{Rubin}). Third, we find that in the semiclassical
limit the eigenstates have a rich structure that reflects the
self-similarity of the sets $K_\pm$ and depends explicitly on the
limiting decay rate. A semiclassical formula is derived for the
weights of the quantum eigenstates in different regions of phase
space. This formula is consistent with the numerical results of
\cite{casati} in the special case when the decay rate is equal to
the classical escape rate.

We illustrate our results for an open quantum baker map. In this
system we have, in addition, that in the semiclassical limit
long-lived right eigenstates have fractal support in momentum space
and are self-similar in position space (and vice-versa for left
eigenstates). Finally, we discuss a system, the Walsh-quantized
baker map, for which exact results corroborate our arguments.

Many recent studies of open quantum systems have focused on maps of
the torus $T$, viewed as a phase space, which is opened by removing
a strip parallel to the momentum direction
\cite{frac2,nonnen,transp1,cross,henning}. A physical motivation is
the ``bounce map" defined for billiards, in which case the opening
corresponds to attaching a perfect lead to the billiard. The
quantization of the classical map $\mathcal{U}$ is a unitary matrix
$U$, acting on a Hilbert space of dimension $N$. For torus maps this
dimension plays the role of an effective Planck's constant,
$\hbar=(2\pi N)^{-1}$, and the semiclassical limit thus corresponds
to $N\to\infty$. If the opening $\mathcal{O}$ occupies a fraction
$M/N$ of the total area in phase space then quantum mechanically the
open map corresponds to a non-unitary matrix $\U=U\Pi$, where $\Pi$
is a projector onto the complement of the opening. The result of
multiplying by $\Pi$ is to set $M$ columns of $U$ equal to zero.
Since the matrix $\U$ is not unitary, we must distinguish between
its left and right eigenstates, \be\label{leftright}
\U|\Psi^R_n\rangle=z_n|\Psi^R_n\rangle, \quad \langle
\Psi_n^L|\U=z_n\langle \Psi_n^L|.\ee We will assume $\langle
\Psi_n^L|\Psi^L_n\rangle=\langle \Psi_n^R|\Psi^R_n\rangle=1$. The
eigenvalues $z_n$ lie inside the unit circle in the complex plane,
$|z_n|^2=e^{-\Gamma_n}\leq 1$ where $\Gamma_n\geq 0$ is the decay
rate. (Note that these are not scattering states, for which
fractality has also been observed \cite{ishio}.)

If $\mathcal{O}_m=\mathcal{U}^m(\mathcal{O})$ denotes the $m$th
image of the opening under the classical map, the forward-trapped
and backward-trapped sets are defined, respectively, as \be
K_+=T\setminus\bigcup_{m=0}^\infty\mathcal{O}_{-m},\quad
K_-=T\setminus\bigcup_{m=1}^\infty\mathcal{O}_{m}.\ee Let us also
define the set of points which fall into the opening after $m$
steps, but not earlier, \be
\mathcal{R}^m_+=\{x\in\mathcal{O}_{-m},\,x\notin\mathcal{O}_{-n},
\,0\leq n<m\}.\ee We also define
$\mathcal{R}^m_-=\{x\in\mathcal{O}_{m},\,x\notin\mathcal{O}_{n},
\,1\leq n<m\}$ for $m>1$ and $\mathcal{R}^1_-=\mathcal{O}_1$. These
sets are related by \be\label{rs}
\mathcal{U}^{-1}(\mathcal{R}_+^m)\setminus
\mathcal{O}=\mathcal{R}_+^{m+1}, \quad
\mathcal{U}(\mathcal{R}_-^m\setminus
\mathcal{O})=\mathcal{R}_-^{m+1},\ee with the convention that
$\mathcal{R}_+^0=\mathcal{O}$.

The Ehrenfest time $\tau_E$ plays an important role in quantum
chaotic transport \cite{larkin,transp1,cross,henning}, essentially
determining the quantum-to-classical crossover. This is the time it
takes for a minimal wave packet to become larger than the opening,
being therefore partially reflected when leaving the system. For
times shorter than $\tau_E$ the quantum evolution of localized
states can be approximated by the classical evolution. For open
chaotic maps $\tau_E=\lambda^{-1}\ln M$, where $M$ is the number of
`open channels' and $\lambda$ is the Lyapunov exponent
\cite{cross,henning}. In the semiclassical limit quantum states that
are essentially supported on some region
$\mathcal{A}\subset\mathcal{R}_+^n$ of phase space will have a
deterministic escape and will thus become eigenstates of $\U^{n+1}$
corresponding to eigenvalues that vanish as $\hbar\to0$. These are
called `short-lived' states \cite{cross,henning}. The estimate that
the fraction of such states is $1-e^{-\tau_E/\tau_D}$, which holds
when the mean dwell time $\tau_D=N/M$ is large, results in the
fractal Weyl law \cite{henning}. Short-lived states are intimately
related to the short-time dynamics and are relatively insensitive to
the classical trapped sets. In what follows we obtain several
results for the semiclassical limit of long-lived eigenstates.

One way to look at the eigenstates is to depict them in phase space
using the Husimi representation, in which one associates with any
eigenstate $|\Psi_n^\xi\rangle$ ($\xi=L,R$) the function
$H_n^\xi(x)=|\langle x|\Psi_n^\xi\rangle|^2$, where $x=(q,p)$ is a
point of phase space and $|x\rangle$ is a coherent state. From
(\ref{leftright}) we have for the Husimi function of a right
eigenstate \be\label{basic} |z_n|^{2m}|\langle
x|\Psi^R_n\rangle|^2=|\langle x| \U^m|\Psi^R_n\rangle|^2.\ee For
times much shorter than the Ehrenfest time the action of the quantum
propagator on coherent states can be approximated by the classical
evolution; in particular we have that as $\hbar\to 0$
\be\label{kill} (\U^\dag)^m|x\rangle\approx 0 \quad{\rm if}\quad
x\in \mathcal{R}_-^m.\ee Therefore, in the semiclassical limit
$H_n^R$ becomes negligible on regions $\mathcal{R}_-^m$ if $|z_n|>0$
is fixed and $m\ll\tau_E$. Since the Ehrenfest time grows as
$\hbar\to0$, the function must concentrate on the backward-trapped
set $K_-$ in this limit, \be H_n^R(x)\approx 0\quad {\rm if} \quad
x\notin K_- \quad(|z_n|>0,\; \hbar\to0).\ee For a finite $\hbar$
there will be some leakage of $|\langle x|\Psi^R_n\rangle|^2$ into
the regions $\mathcal{O}_m$, mostly for large $m$. This deviation
should be more pronounced for states with small lifetime. The
reasoning for long-lived left eigenstates is completely analogous.
In the semiclassical limit they concentrate on the forward-trapped
set $K_+$.

For finite $\hbar$ the Husimi functions are not supported on truly
fractal sets, but as $\hbar\to0$ finer classical structures are
revealed. Continuous time Hamiltonian systems can be treated in much
the same way. One just replaces $|z_n|^{2m}$ by $e^{-\Gamma_nT}$ and
the essence of the argument remains. We thus expect Husimi functions
that live on fractal sets in generic open chaotic systems in the
semiclassical limit.

Let us investigate the weight of the eigenstates on different
regions of phase space. Let $\pi_0$ be the projector onto the
opening. Our non-unitary propagator then satisfies
$\U^\dag\U=1-\pi_0$. Hence we find that the weight of the right
eigenstates in the opening, \be \label{opening} \langle
\Psi_n^R|\pi_0|\Psi_n^R\rangle=1-|z_n|^2,\ee increases with
increasing decay rate. Let $\pi_m$, $m\geq 1$, be the projector onto
$\mathcal{R}_+^m$ (this is an example of the general class of
projectors introduced in \cite{raul}). For short times the
semiclassical approximation gives
$\U^\dag\pi_{m}\U\approx\pi_{m+1}$, and thus we obtain the
remarkably simple relation \be\label{weight} \langle
\Psi_n^R|\pi_{m}|\Psi_n^R\rangle\approx|z_n|^{2m}(1-|z_n|^2).\ee
Notice that this vanishes for any fixed $m$ as $|z_n|\to 1$, thus if
the quantum decay rate approaches zero as $\hbar\to0$ the
corresponding eigenstate becomes localized on the invariant set
$K_0$.

It is interesting to note that the distribution of the eigenstates
on phase space depends explicitly on the corresponding decay rate,
which we hold fixed as $\hbar\to0$. This is an important difference
with respect to the case of unitary evolution, when the
semiclassical limit is the same for almost all sequences of
eigenstates $\{\psi(\hbar)\}_{\hbar\to 0}$. We also note that the
areas of the regions $\mathcal{R}_+^m$ are proportional to
$e^{-m\gamma_c}$, where $\gamma_c$ is the classical escape rate.
Therefore the prediction in \cite{casati} that a right eigenstate of
an open quantum map with $\Gamma_n\approx \gamma_c$ should be
constant (up to quantum fluctuations) over the classical set $K_-$
is consistent with our expression.

\begin{figure}[t]
\includegraphics[scale=0.42]{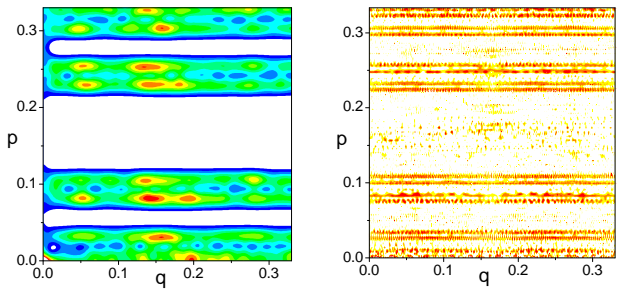}
\caption{(color online) Left panel: The average of the Husimi
functions of the $100$ longest-lived right eigenstates of the baker
map, for $N=3^7$ (intensity increases from blue to red). Right
panel: the corresponding Wigner function average (in the white
regions the function is non-positive). Note that the $\hbar$-scale
is $1/N\approx0.005$.}
\end{figure}

We now illustrate our results in a paradigmatic model of quantum
chaos, the baker map, which is particularly convenient because of
the relatively simple structure of its trapped sets. Its triadic
form is defined as \be
\mathcal{U}(q,p)=\begin{cases}(3q,\frac{p}{3}) &\text{if } 0\leq q<\frac{1}{3}, \\
(3q-1,\frac{p+1}{3}) &\text{if } \frac{1}{3}\leq
q<\frac{2}{3},\\(3q-2,\frac{p+2}{3}) &\text{if } \frac{2}{3}\leq
q<1.\end{cases}\ee This system is uniformly hyperbolic, with
Lyapounov exponent $\lambda=\ln 3$. The stable and unstable
manifolds are parallel to the momentum and position axes,
respectively. Its quantization is given by \cite{baker} \be
U_N=F_N^{-1}{\rm diag}(F_{N/3},F_{N/3},F_{N/3}),\ee where the
subscript denotes the dimension of the matrix. $F_N$ is a modified
Fourier transform, $(F_N)_{nm}=\frac{1}{\sqrt{N}}e^{-\frac{2\pi
i}{N}(n+1/2)(m+1/2)},$ with the integers $n,m$ running from $0$ to
$N-1$. The ``$1/2$" factors are necessary in order to impose parity
on the eigenstates. This form of the matrix $F_N$ leads to
antiperiodic boundary conditions.

An open version of this system was introduced in
\cite{frac2,nonnen}, in the context of the fractal Weyl law, where
the opening was taken as the middle vertical strip (see also
\cite{raul1} where a closely related map was studied). To find the
forward trapped set $K_+$ one may consider the backward propagation
of this strip, and it is not difficult to see that the only points
that remain are those belonging to ${\rm Can}\times[0,1)$, where
`${\rm Can}$' denotes the usual middle-third fractal Cantor set. The
backward-trapped set $K_-$ is $[0,1)\times{\rm Can}$. Details of
this construction can be found in \cite{frac2,nonnen}. The open
quantum system is obtained by setting the middle third of the
columns of $U_N$, corresponding to the strip, equal to zero. The
kinematics is such that the position representation of right
eigenstates is equal to the momentum representation of left
eigenstates.

In Fig.1 we plot the average of $H_n^R(q,p)$ over the $100$
longest-lived states, for the case $N=3^7$. We see that it is
reasonably concentrated on the backward-trapped set, although it is
not able to resolve this set on the finest scales. More details of
$K_-$ should be revealed for larger values of $N$. The right panel
of Fig.1 shows the averaged Wigner function, and we see that it
resolves $K_-$ with considerably greater accuracy.

\begin{figure}[t]
\includegraphics[scale=0.44]{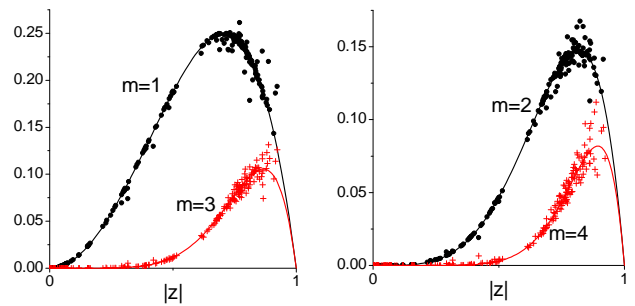}
\caption{(color online) The weight $\langle
\Psi_n^R|\pi_{m}|\Psi_n^R\rangle$ of right eigenstates of the baker
map on the regions $\mathcal{R}_+^m$, for $N=3^6$. The solid line is
the semiclassical approximation \eqref{weight}.}
\end{figure}

In Fig.2 we plot the weight of right eigenstates on the regions
$\mathcal{R}_+^m$, for $N=3^6$. The solid lines represent the
semiclassical approximation \eqref{weight}. We see that the
agreement is rather good, particularly for small eigenvalue modulus
and small values of $m$.

Let us now turn our attention to wave functions in position/momentum
space. For the baker map, the projection of $K_-$ onto momentum
space is just the Cantor set, and the momentum representation of any
long-lived right eigenstate must be supported on this fractal dust.
In Fig.3 we plot $|\Psi^R(p)|^2$, averaged over the 25 longest-lived
states, for $N=3^7$. The fractal nature of the support, within the
$\frac{1}{N}$ resolution set by the uncertainty principle, is
evident from the magnification by a factor of $3$. In Fig.4 we plot
$|\langle q|\Psi^R\rangle|^2$ averaged over a few right eigenstates,
again for $N=3^7$. Panels $a$ and $b$ correspond to a relatively
large decay rate, while $c$ and $d$ correspond to a smaller one.
Within the available resolution both functions are remarkably
self-similar, as can be appreciated from the magnifications. This
observation is consistent with our theory (cf. (\ref{weight})).

\begin{figure}[t]
\includegraphics[scale=0.43]{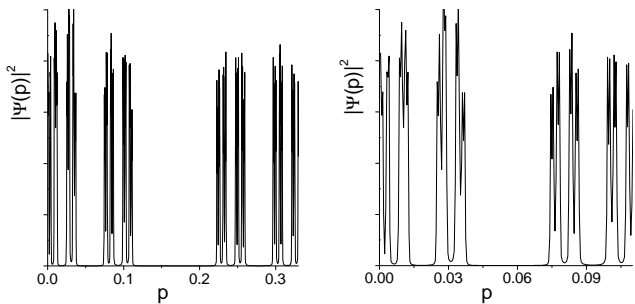}
\caption{Probability density in momentum space for right
eigenstates, $|\langle p|\Psi^R\rangle|^2$, averaged over the $20$
longest-lived states (eigenvalue modulus ranging from $0.90$ to
$0.83$). This function is approximately supported on the Cantor set,
as can be seen from a magnification by a factor of $3$ (right
panel). $N=3^7$ and so the $\hbar$-scale is $1/N\approx0.005$.}
\end{figure}

\begin{figure}[b]
\includegraphics[scale=0.44]{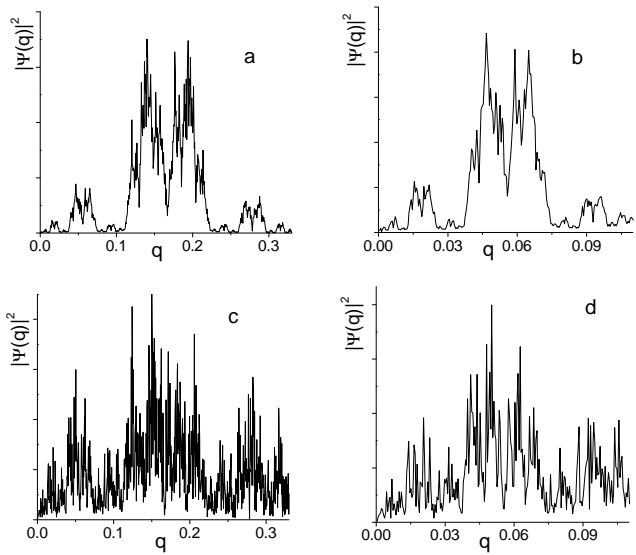}
\caption{Probability density in position space for right
eigenstates, $|\langle q|\Psi^R\rangle|^2$, averaged over states
with similar decay rates. Panels $b$ and $d$ are magnifications of
$a$ and $c$. The average eigenvalue modulus is approximately $0.4$
for $a$-$b$ and $0.7$ for $c$-$d$. In both cases the self-similarity
is striking. $N=3^7$.}
\end{figure}

Nonnenmacher and Zworski have recently introduced a simplified
version of the open quantum baker map, based on the `Walsh-Fourier'
transform, that can be solved explicitly \cite{nonnen}. For this
system several of the results presented here are exact and can be
rigorously proven. One particular feature is that the difference
between short-lived and long-lived states is well defined. The
former all have exactly null eigenvalue, and the phase space
representation of the latter vanishes outside the finitely resolved
relevant trapped set. Equations (\ref{kill}) and (\ref{weight}) are
exact for this system. We have also shown that self-similar
properties analogous to those illustrated by Fig.4 can be
established rigorously and that --after either a spatial average or
a spectral average-- the probability density of long-lived right
eigenstates in the momentum representation becomes constant on the
Cantor set as $N\to\infty$ (for the standard quantization this is
suggested by Fig.3). A detailed account will be presented elsewhere
\cite{us}.

Summarizing, we have shown that in the semiclassical limit resonance
eigenstates of open chaotic systems concentrate on classical fractal
trapped sets. We have derived a formula for their distribution on
different areas of phase space that depends explicitly on the
quantum decay rate. For the baker map we have also found that the
position and momentum representations of wave functions exhibit
self-similarity.

This research was supported by EPSRC and the Royal Society. One of
the authors (M.N.) also gratefully acknowledges financial support
from CAPES.


\begin{thebibliography}{99}

\bibitem{weyl} M. Dimassi and J. Sj\"ostrand, {\it Spectral
asymptotics in the semiclassical limit}, (Cambridge Univ. Press,
Cambridge, 1999).

\bibitem{berry} M.V. Berry, J. Phys. A {\bf 10}, 2083 (1977).

\bibitem{ergod} A.I. Shnirelman, Usp. Mat. Nauk. {\bf 29} (6), 181
(1974).

\bibitem{frac1} J. Sj\"ostrand, Duke Math.J. {\bf 60}, 1 (1990); M.
Zworski, Not. Am. Math. Soc. {\bf 46}, 319 (1999)

\bibitem{frac2} J. Sj\"ostrand and M. Zworski, {\rm math.SP}/0506307.

\bibitem{frac3} K.K. Lin, J. Comput. Phys. {\bf 176}, 295 (2002); K.K. Lin
and M. Zworski, Chem. Phys. Lett. {\bf 355}, 201 (2002); W.T. Lu, S.
Sridhar and M. Zworski, Phys. Rev. Lett. {\bf 91}, 154101 (2003).

\bibitem{Yan} Y.V. Fyodorov and H-J. Sommers, J. Math. Phys. {\bf 38}, 1918 (1997);
J. Phys. A {\bf 36}, 3303 (2003).

\bibitem{phys} O.I. Tolstikhin, V.N. Ostrovsky and H. Nakamura,
Phys. Rev. Lett {\bf 79}, 2026 (1997); Phys. Rev. A {\bf 58}, 2077
(1998); G.V. Sitnikov and O.I. Tolstikhin, {\it ibid} {\bf 67},
032714 (2003).

\bibitem{chem} B.K. Kendrik {\it et al}, Phys. Rev. Lett {\bf 84},
4325 (2000); R.T. Skodje {\it et al}, J. Chem. Phys. {\bf 112}, 4536
(2000); G.C. Schatz, Science {\bf 288}, 1599 (2000);

\bibitem{casati} G. Casati, G. Maspero and D.L. Shepelyansky,
Physica D {\bf 131}, 311 (1999).

\bibitem{Rubin} S. Nonnenmacher and M. Rubin, in preparation.

\bibitem{nonnen} S. Nonnenmacher and M. Zworski, J. Phys. A {\bf 38}, 10683
(2005). See also {\rm math-ph}/0505034.

\bibitem{transp1} P.G. Silvestrov, M.C. Goorden and
C.W.J. Beenakker, Phys. Rev. B {\bf 67}, 241301 (2003); J. Tworzydlo
{\it et al.}, {\it ibid} {\bf 68}, 115313 (2003); Ph. Jacquod and
E.V. Sukhorukov, Phys. Rev. Lett. {\bf 92}, 116801 (2004); S. Rahav
and P.W. Brouwer, {\it ibid} {\bf 95}, 056806 (2005); {\it ibid}
 {\bf 96}, 196804 (2006); Ph. Jacquod and R.S. Whitney, {\rm
cond-mat}/0512662.

\bibitem{cross} H. Schomerus and Ph. Jacquod, J. Phys. A {\bf
38}, 10663 (2005).

\bibitem{henning} H. Schomerus and J. Tworzydlo, Phys. Rev. Lett. {\bf
93}, 154102 (2004).

\bibitem{ishio} H. Ishio and J.P. Keating, J. Phys. A {\bf 37},
L217 (2004).

\bibitem{larkin} I. Aleiner and A. Larkin, Phys. Rev. B {\bf
54}, 14423 (1996); S. Oberholzer, E.V. Sukhorukov and C.
Schonenberger, Nature {\bf 415}, 765 (2002).

\bibitem{raul} M. Saraceno and A. Voros, Physica D {\bf 79}, 206 (1994);
R.O. Vallejos and M. Saraceno, J. Phys. A {\bf 32}, 7273 (1999).

\bibitem{baker} N.L. Balazs and A. Voros, Ann. Phys. (NY)
{\bf 190}, 1 (1989); M. Saraceno, {\it ibid} {\bf 199}, 37 (1990).

\bibitem{raul1} M. Saraceno and R.O. Vallejos, Chaos {\bf 6}, 193
(1996).

\bibitem{us} J.P. Keating {\it et al.} (to be published).

\end{thebibliography}
\end{document}